\documentclass{article}
\usepackage{spconf,amsmath,graphicx}
\usepackage{amssymb}
\usepackage{amsthm}
\usepackage{color}
\usepackage{tabularx}
\usepackage{multirow}
\usepackage{booktabs}
\usepackage{bm}
\usepackage{qtree}
\usepackage{tikz}
\usepackage{mathtools}
\usepackage[colorlinks]{hyperref}
\usepackage{caption}

\newtheorem{proposition}{Proposition}
\newtheorem{definition}{Definition}

\title{Local, global and scale-dependent node roles}
\name{Michael Scholkemper, Michael T. Schaub\thanks{We acknowledge partial funding from Ministry of Culture and Science of North Rhine-Westphalia (NRW Rückkehrprogramm) and the Excellence Strategy of the Federal Government and the Länder.}}
\address{Department of Computer Science, RWTH Aachen University, Germany}
\begin{document}
\ninept
\maketitle
\begin{abstract}
    This paper re-examines the concept of node equivalences like \emph{structural equivalence} or \emph{automorphic equivalence}, which have originally emerged in social network analysis to characterize the role an actor plays within a social system, but have since then been of independent interest for graph-based learning tasks.
    Traditionally, such exact node equivalences have been defined either in terms of the one-hop neighborhood of a node, or in terms of the global graph structure.
    Here we formalize exact node roles with a scale-parameter, describing up to what distance the ego network of a node should be considered when assigning node roles --- motivated by the idea that there can be local roles of a node that should not be determined by nodes arbitrarily far away in the network.
    We present numerical experiments that show how already ``shallow'' roles of depth $3$ or $4$ carry sufficient information to perform node classification tasks with high accuracy.
    These findings corroborate the success of recent graph-learning approaches that compute approximate node roles in terms of embeddings, by nonlinearly aggregating node features in an (un)supervised manner over relatively small neighborhood sizes.
    Indeed, based on our ideas we can construct a shallow classifier achieving on par results with recent graph neural network architectures.
\end{abstract}
\section{Introduction}

\label{sec:intro}

Networks have become a powerful abstraction to understand a range of complex systems \cite{newman2018networks, strogatz2001exploring}.
To comprehend such networks we often seek patterns in their connections, e.g., densely-knit clusters or core-periphery structure, which would enable a simpler or faster analysis of such systems.
The concept of node roles or node equivalences, originating in social network analysis \cite{knoke2019social}
is another of those patterns used to simplify complex networks. 
The questions underpinning the detection of node roles are i) what nodes serve what function within the network? and ii) which nodes are similar in functionality?

The intricacy of node role extraction derives from the lack of a clear definition of the role of a node in mathematical terms.
Traditional approaches, put forward in the context of social network analysis \cite{brandes2005network}
consider \emph{exact node equivalences}, based on structural symmetries within the graph structure.
The earliest notion is that of \textit{structural equivalence} \cite{lorrain1971structural}, which assigns the same role to two nodes if they are adjacent to the exact same nodes.
Another definition is that of  \textit{automorphic equivalence} \cite{everett1994regular}, which states that nodes are equivalent if they belong to the same automorphism orbits.
Closely related is the idea of \textit{regular equivalent} nodes \cite{white1983graph}, defined recursively as nodes that are adjacent to equivalent nodes.

Interestingly, essentially all of these approaches determine the role of the node either purely on the direct neighborhood of a node, or in terms of the ``global neighborhood'', i.e., the relative position of a node within the whole graph.
Motivated by this observation, here we re-consider such exact node equivalences and introduce a scale parameter for the definition of exact node equivalences, and discuss algorithms that are able to find the thus defined (local) node roles.
The overall idea here is that rather than having a single, fixed scale for defining a node role, there may be different relevant scales for node roles, depending on the problem at hand --- indeed one may argue that, e.g., in a social network context the role of a node should not be dependent on nodes potentially arbitrary far away.

\textbf{Related literature}
Apart from the above mentioned ideas of \emph{exact} node equivalences, there exists a large number of works on role extraction, which focus on finding nodes with \emph{similar} roles (though not identical).
Many of these methods are based on computing feature vectors or embeddings of nodes.
Based on these embedding we can then calculate pair-wise similarities between nodes and cluster them into groups. 
The overview article~\cite{rossi2020proximity} puts forward three categories:
First, graphlet-based approaches \cite{prvzulj2007biological, rossi2018estimation, liu2020learning} use the number of graph homomorphisms of small structures to create node embeddings. 
This retrieves extensive, local information such as the number of triangles a node is part of. 
Second, walk-based approaches \cite{ahmed2019role2vec, cooper2010role} embed nodes based on certain statistics of random walks starting at each node. 
Finally, matrix-factorization-based approaches \cite{henderson2011rolx, jin2019smart} find a rank-$r$ approximation of a node feature matrix ($F\approx MG$). Then, the left side multiplicand $M \in \mathbb{R}^{|V| \times r}$ of this factorization is used as a soft assignment of the nodes to $r$ clusters. 

Jin et al. \cite{jin2021towards} provide a comparison of many of such node embedding techniques in terms of their ability to capture exact node roles such as structural, automorphic and regular node equivalence. 
Detailed overviews of (exact) role extraction and its links to related topics such as block-modelling are also given in \cite{browet2014algorithms, cason2012role}. 
The idea of node roles is also similar to community detection \cite{fortunato2010community}.
However, whereas the locality of connections is often a central factor for community detection, this is less of a consideration for node roles.
Indeed, nodes that are far apart or are even part of different connected components, can have the same role~\cite{rossi2020proximity}.

\textbf{Contributions and outline.} 
We provide a fresh perspective on the definition of exact node roles which has certain parallels in recent developments in graph-based machine learning techniques such as graph neural networks. 
First, we formalize exact node roles with a scale parameter, describing up to what distance the graph structure surrounding a node should be considered when assigning node roles. 
Second, we re-frame the well-known Weisfeiler Lehman (WL) algorithm, and show how it can be interpreted in terms of a relaxation of computing such local node roles.
We then provide a different algorithm that solves the scale-dependent node embedding, using an alternative problem relaxation, which may be interpreted as a dual of the WL algorithm.
Finally, we show that \textit{shallow} roles provide enough information to achieve high performance on node classification tasks, and compare these results based on local node roles with the performance of recent graph neural network architectures.

\section{Notation and Background}
\label{sec:background}

\textbf{Graphs.}
An undirected \textit{graph} $G$ consists of a vertex-set $V$ and edge-set $E \subseteq \{\{u,v\} | u,v\in V\}$ describing relations between the vertices.
Given a subset $V' \subseteq V$, the \textit{graph induced by} $V'$ is defined as the graph with vertex set $ V'$ and edge set $E' = \{\{u,v\} \in E | u,v\in V'\}$. 
For a vertex $v$, we define its \textit{neighbourhood} as the set $N(v) = \{x | \{v, x\} \in E\}$.
The \textit{$k$-hop-neighbourhood} $N^k(v)$ is the set of nodes reachable from $v$ in at most $k$ steps. 

\noindent\textbf{Colorings and refinements.}
We define a graph \textit{coloring} as a function $c : V \rightarrow \{1,\ldots, \mathcal{C}\}$ which maps each node to one out of $\mathcal{C}\in\mathbb{N}$ many colors.
The classes of a coloring are the node sets associated to the same color $C_i = \{v \in V \ | \ c(v) = i\}$. 
The partition associated to these classes induces an equivalence relation $\sim_c$ among the vertices. 
We say a coloring $c$ \textit{refines} another coloring $c'$, written $c\sqsubseteq c'$, if for all node $u, v \in V$ a different color assignment $c'(u) \neq c'(v)$ under coloring $c'$ implies a different assignment $c(u) \neq c(v)$ under coloring $c$. 
Hence, the partition induced by the color-classes of $c$ is a subpartition of the partition induced by $c'$. 
Two colorings $c, c'$ are \textit{equivalent}, written $c \equiv c'$, if they refine each other ($c \sqsubseteq c'$ and $c'\sqsubseteq c$).
This implies that  $\sim_c$ and $\sim_{c'}$ define the same relation. 
A \textit{refinement} is an iterative algorithm, that produces at each iteration $t$ a coloring $c^t$ that refines the previous coloring, i.e., $c^{t+1} \sqsubseteq c^t$.

\noindent\textbf{Isomorphism and orbits.}
An isomorphism between two graphs $G, G'$ is function that maps all vertices from one graph to the other graph, while preserving adjacency relationships and coloring.
Formally, given colored graphs $(G, c), (G', c')$, an isomorphism is a bijection $\pi:~V(G) \rightarrow V(G')$ such that (i) $\{u,v\} \in E(G) \Leftrightarrow \{\pi(u), \pi(v)\} \in E(G')$, and (ii) $c(v) = c(\pi(v))$ for all vertices. 
If such an isomorphism exists between graphs $G$ and $G'$, we say that these graphs are \textit{isomorphic} $G \cong G'$. 
An \textit{automorphism} is an isomorphism from a graph to itself $\pi' : G \rightarrow G$. 
The \textit{orbit} $\text{orb}_G(v)$ of a vertex $v$ is the set of all vertices $u$ for which there exists an automorphism $\pi'$ such that $\pi'(v)=u$.

\noindent\textbf{Unravellings.} 
Let $(G, c)$ be a colored graph and $v \in V$. 
The \textit{node-unidentified unravelling} $\mathcal{U}^d(v)$ of depth $d$ rooted at node $v$ is the tree defined as follows.
The vertex set of $\mathcal{U}^d(v)$ is the set of walks of length at most $d$ starting at $v$. 
Furthermore two nodes $w_1 =(v, x_1, ..., x_n)$ and $w_2= (v, y_1, ..., y_{n-1})$ are connected if $x_i = y_i$ for $1 \leq i \leq n-1$,
i.e., if the walk corresponding to $w_1$ is simply an extension by one node of the walk corresponding to $w_2$.
This definition induces a natural equivalence relation between two nodes $u \sim_{\mathcal{U}}^d v$, where nodes are equivalent if and only if $\mathcal{U}^d(u) \cong \mathcal{U}^d(v)$.

The \textit{node-identified unravelling} $\mathcal{I}^d(v)$ of depth $d$ rooted at node $v$ is defined analogously as above, with one addition: Each node $(v, x_1, ..., x_n)$ is identified by the vertex $x_n$ at the end of the walk, i.e. $\text{id}((v, x_1, ..., x_n)) = x_n$. 
For two nodes $u,v$ to be equivalent  $u \sim_{\mathcal{I}} v$, there must exist an isomorphism $\sigma$ certifying $\mathcal{U}^d(u) \cong \mathcal{U}^d(v)$ such that $\pi : \text{id}(x) \rightarrow \text{id}(\sigma(x))$ is a well-defined bijection. 
In other words, all nodes in $\mathcal{U}^d(u)$ with the same id $i$ are consistently mapped onto nodes with the same id $j$ in $\mathcal{U}^d(v)$ by $\sigma$ (but it is not required that $i = j$). 
The motivation underpinning this definition is that a consistent isomorphism $\sigma$ between the unravellings induces a local isomorphism $\pi$ between the graphs (see Prop.~\ref{prop:automorphism_orbits}, Fig.~\ref{figure_higher_distinguishing_power}). 

\begin{figure}[th]
	
	\centering	
	\begin{minipage}{.15 \textwidth}
	    \tiny
		\centering
		\caption*{$G$}
		\vspace{-0.3cm}
		\begin{tikzpicture}[->,>=stealth,shorten >=1pt,auto,node distance=0.5cm,
		thick,main node/.style={circle,draw,minimum size=0.3cm,inner sep=0pt}]
		
		\node[main node, blue] (1) {$1$};
		\node[main node, red] (2) [above right of=1]  {$2$};
		\node[main node, green] (3) [below right of=1] {$3$};
		\node[main node, orange] (4) [right of=2] {$4$};
		\node[main node, brown] (5) [below right of=4] {$5$};
		\node[main node, cyan] (6) [below left of=5] {$6$};

		\path[-]
		(1) edge node {} (2)
		edge node {} (3)
		(3) edge node {} (1)
		edge node {} (6)
		(6) edge node {} (3)
		edge node {} (5)
		(5) edge node {} (6)
		edge node {} (4)
		(4) edge node {} (5)
		edge node {} (2)
		(2) edge node {} (4)
		edge node {} (1)
		
		;
		\end{tikzpicture}
		
	\end{minipage}
	\begin{minipage}{.24 \textwidth}
	    \centering
	    \caption*{$\mathcal{U}^3(1)$}
	    \vspace{-0.3cm}
	  \begin{tikzpicture}[level distance=0.25cm,
		level 1/.style={sibling distance=2cm},
		level 2/.style={sibling distance=0.95cm},
		level 3/.style={sibling distance=0.4cm}]
		
		\tikzstyle{every node}=[circle,draw, fill,minimum size=0.1cm,inner sep=0pt]
		\node [black] (Root) {}
		child {
			node [black] {}
			child { 
				node [black] {}
				child { node [black] {} }
				child { node [black] {} } 
			}
			child { 
				node [black] {}
				child { node [black] {} }
				child { node [black] {} } 
			}
		}
		child {
			node [black] {}
			child { 
				node [black] {}
				child { node [black] {} }
				child { node [black] {} } 
			}
			child { 
				node [black] {}
				child { node [black] {} }
				child { node [black] {} } 
			}
		};
		\end{tikzpicture}
	\end{minipage}
	
	\begin{minipage}{.15 \textwidth}
		\centering
		\tiny
	    \begin{tikzpicture}[->,>=stealth,shorten >=1pt,auto,node distance=0.5cm,
		thick,main node/.style={circle,draw,minimum size=0.3cm,inner sep=0pt]}]
		
		\node[main node, blue] (1) {$1$};
		\node[main node, orange] (2) [above right of=1]  {$4$};
		\node[main node, brown] (3) [below right of=1] {$5$};
		\node[main node, red] (4) [right of=2] {$2$};
		\node[main node, green] (5) [below right of=4] {$3$};
		\node[main node, cyan] (6) [below left of=5] {$6$};

		\path[-]
		(1) edge node {} (2)
		edge node {} (3)
		(3) edge node {} (1)
		edge node {} (2)
		(6) edge node {} (4)
		edge node {} (5)
		(5) edge node {} (6)
		edge node {} (4)
		(4) edge node {} (5)
		edge node {} (6)
		(2) edge node {} (3)
		edge node {} (1)
		
		;
		\end{tikzpicture}
		
	\end{minipage}
	\begin{minipage}{.24 \textwidth}
	    \centering
	    \tiny
	  \begin{tikzpicture}[level distance=0.25cm,
		level 1/.style={sibling distance=2cm},
		level 2/.style={sibling distance=0.95cm},
		level 3/.style={sibling distance=0.4cm}]
		
		\tikzstyle{every node}=[circle,draw, fill,minimum size=0.1cm,inner sep=0pt]
		\node [black] (Root) {}
		child {
			node [black] {}
			child { 
				node [black] {}
				child { node [black] {} }
				child { node [black] {} } 
			}
			child { 
				node [black] {}
				child { node [black] {} }
				child { node [black] {} } 
			}
		}
		child {
			node [black] {}
			child { 
				node [black] {}
				child { node [black] {} }
				child { node [black] {} } 
			}
			child { 
				node [black] {}
				child { node [black] {} }
				child { node [black] {} } 
			}
		};
		\end{tikzpicture}
	\end{minipage}
	
	\vspace{0.2cm}
	
	\begin{minipage}{.15 \textwidth}
	    \scriptsize
	    \centering
	    \caption*{$\text{SNP}^3(1)$}
	    \vspace{-0.3cm}
	    \begin{tikzpicture}[level distance=0.25cm,
		level 1/.style={sibling distance=2cm},
		level 2/.style={sibling distance=0.95cm},
		level 3/.style={sibling distance=0.4cm}]
		
		\begin{scope}[every node/.style={right}]
		\path (Root    -| Root-2-2) ++(0,0) node [align=left] {$(0,0,0,0,0,\textcolor{blue}{1})$};
		\path (Root-1  -| Root-2-2) ++(0,0) node [align=left] {$(0,0,0,\textcolor{red}{1},\textcolor{green}{1},0)$};
		\path (Root-1-1-| Root-2-2) ++(0,0) node [align=left] {$(0,\textcolor{orange}{1},\textcolor{cyan}{1},0,0,\textcolor{blue}{2})$};
		\path (Root-1-1-1-| Root-2-2) ++(0,0) node [align=left] {$(\textcolor{brown}{2},0,0,\textcolor{red}{3},\textcolor{green}{3},0)$};
		\end{scope}

		\end{tikzpicture}
	\end{minipage}
	\begin{minipage}{.24 \textwidth}
	\centering
	\scriptsize
    \caption*{$\mathcal{I}^3(1)$}
    \vspace{-0.3cm}
	  \begin{tikzpicture}[level distance=0.25cm,
		level 1/.style={sibling distance=2cm},
		level 2/.style={sibling distance=0.95cm},
		level 3/.style={sibling distance=0.4cm}]
		
		\tikzstyle{every node}=[circle,minimum size=0.2cm,inner sep=0pt]
		\node [blue] (Root) {1}
		child {
			node [red] {2}
			child { 
				node [blue] {1}
				child { node [red] {2} }
				child { node [green] {3} } 
			}
			child { 
				node [orange] {4}
				child { node [red] {2} }
				child { node [brown] {5} } 
			}
		}
		child {
			node [green] {3}
			child { 
				node [blue] {1}
				child { node [red] {2} }
				child { node [green] {3} } 
			}
			child { 
				node [cyan] {6}
				child { node [green] {3} }
				child { node [brown] {5} } 
			}
		};
		\end{tikzpicture}
	\end{minipage}

	\begin{minipage}{.15 \textwidth}
	    \centering
	    \scriptsize
	    \begin{tikzpicture}[level distance=0.25cm,
		level 1/.style={sibling distance=2cm},
		level 2/.style={sibling distance=0.95cm},
		level 3/.style={sibling distance=0.4cm}]
		\begin{scope}[every node/.style={right}]
		\path (Root    -| Root-2-2) ++(0,0) node [align=left] {$(0,0,0,0,0,\textcolor{blue}{1})$};
		\path (Root-1  -| Root-2-2) ++(0,0) node [align=left] {$(0,0,0,\textcolor{orange}{1},\textcolor{brown}{1},0)$};
		\path (Root-1-1-| Root-2-2) ++(0,0) node [align=left] {$(0,0,0,\textcolor{orange}{1},\textcolor{brown}{1},\textcolor{blue}{2})$};
		\path (Root-1-1-1-| Root-2-2) ++(0,0) node [align=left] {$(0,0,0,\textcolor{orange}{3},\textcolor{brown}{3},\textcolor{blue}{2})$};
		\end{scope}
		\normalsize
		\end{tikzpicture}
	\end{minipage}
	\begin{minipage}{.24 \textwidth}
	\centering
	\scriptsize
		\begin{tikzpicture}[level distance=0.25cm,
		level 1/.style={sibling distance=2cm},
		level 2/.style={sibling distance=0.95cm},
		level 3/.style={sibling distance=0.4cm}]
		\tikzstyle{every node}=[circle,minimum size=0.1cm,inner sep=0pt]
		\node [blue] (Root) {1}
		child {
			node [orange] {4}
			child { 
				node [blue] {1}
				child { node [orange] {4} }
				child { node [brown] {5} } 
			}
			child { 
				node [brown] {5}
				child { node [blue] {1} }
				child { node [orange] {4} } 
			}
		}
		child {
			node [brown] {5}
			child { 
				node [blue] {1}
				child { node [orange] {4} }
				child { node [brown] {5} } 
			}
			child { 
				node [orange] {4}
				child { node [blue] {1} }
				child { node [brown] {5} } 
			}
		};
		\end{tikzpicture}
	\end{minipage}
	\caption{\textbf{Illustration of different $d$-role definitions.} The figure shows two graphs, the identified and unidentified unravellings as well as the SNP-embeddings of node 1 in the respective graph. The $0$- and $1$-roles of the nodes in the graphs are the same, but for $d \geq 2$, the $d$-roles are different. There are 5 distinct ids in the upper unravelling, but only 3 distinct ids in the lower one, thus there exists no bijection that is consistent with respect to the ids. This is reflected in the SNP-embedding whose roles are also different for $d \geq 2$. The $d$-wl-roles, however, are the same for all nodes and all $d$. (Colors are only used for visual clarity, they are interchangeable.)}
	\label{figure_higher_distinguishing_power}
\end{figure}

\section{Local, global and scale dependent roles}
\label{sec:noderoles}

Though assigning roles to a node in a network is an intuitively simple idea, there is an inherent tension in the definitions of roles:
on the one hand we want to identify roles that help us to comprehend the system on a global level, on the other hand the semantics of the data encoded in the network is often much more local.

To illustrate this, consider a social network. 
Intuitively, the direct neighbourhood of a node defines its sphere of influence and thus should be considered when analysing which role a node plays. 
Applying this argument recursively, the same is true for the second hop neighbourhood of a node.
Following this train of thought, we may argue that automorphism orbits define nodes roles in the most natural way: these are the finest possible role definitions which are isomorphism preserving and thus independent of the ordering of the node labels.
Yet, with every hop, the relative influence of a node within a social network is bound to decrease.
Automorphic equivalence may thus lead to undesirable node assignments, as any minor asymmetry in the graph far away from a node could influence its role. 
We may thus prefer a more local definition of node roles.

However, a global view on node roles may indeed be desired in other scenarios. 
For instance, in molecular biology we may encode a protein as a graph of amino acids and their interactions.
However, the overall protein structure can be shaped by even far apart amino acids, local changes in the amino acids can influence the functionality of reaction sites and the whole protein. 

An inherent scale is also present in most approaches to extract approximate node roles~\cite{rossi2020proximity,cason2012role}.
For computational reasons, most methods calculate feature vectors of local node statistics, which are then aggregated via clustering.
For example, graphlet density calculation \cite{prvzulj2007biological,rossi2018estimation}, or random walk based statistics \cite{ahmed2019role2vec, cooper2010role} result in highly local features for every node, which are then used to compute approximate node roles.
However, the scale of the extracted features is thus fixed and not adaptable to the problem at hand. 
In fact in some cases, there may be multiple sensible node-role assignments, depending on the scale of the problem one is interested in.

We therefore propose a formal definition of node roles that combines these demands through a scale-parameter. 
\begin{definition}
The \textit{$d$-depth node role} ($d$-role) is given by the structural equivalence of the node-identified unravellings. Equivalently, two nodes $u, v$ have the same $d$-roles if and only if $u \sim^d_{\mathcal{I}}v$.  
\end{definition}

Thus, the $d$-role of a node is local for small $d$, since only nodes in the $d$-hop neighbourhood are used to obtain the roles. 
In contrast, the $d$-role can also include more global features, if $d$ is chosen sufficiently large. 
This mirrors developments in community detection \cite{fortunato2010community}, the analysis of mixing patterns \cite{peel2018multiscale}, or in centrality measures such as Katz centrality \cite{katz1953new}, the scale of interest can often be adjusted via a parameter.

Our first shows that for large $d$, our node role definition  coincides with automorphic equivalence.

\begin{proposition}
    \label{prop:automorphism_orbits}
    Let $G_1, G_2$ be two (colored) graphs of the same size and $u \in V(G_1), v \in V(G_2)$. For $d = \max(|V(G_1)|, |V(G_2)|)$, it holds that $u\sim_{\mathcal{I}}^d v$ if and only if $u \in \text{orb}_{G_1\cup G_2}(v)$.
\end{proposition}

\begin{proof}
For the backward direction, suppose $u \in \text{orb}(v)$ and let $\pi$ be the automorphism certifying this. 
Then the permutation
$$\sigma((v, x_1, ..., x_n)) \coloneqq (\pi(v), \pi(x_1), ..., \pi(x_n))$$ 
proves that $\mathcal{U}^d(u) \cong \mathcal{U}^d(v)$ and $\text{id}(x) \rightarrow \text{id}(\sigma(x)) = \pi(\text{id}(x))$ is well-defined and bijective.

For the other direction, suppose there exists a $\sigma$ that proves $\mathcal{U}^d(v) \cong \mathcal{U}^d(u)$, such that $\pi : \text{id}(x) \rightarrow \text{id}(\sigma(x))$ is well-defined and bijective. 
Let $\Tilde{E}(v) = E(G_1[N^d(v)])$, i.e. the set of all edges reachable from $v$. Consider the edge $\{x_1, x_2\} \in \Tilde{E}(v)$. 
Then there exist both a walk $w_1$ starting at $v$ and ending in $x_1$ as well as a walk $w_2$ that is the same as $w_1$ but extended by $x_2$. $w_1$ and $w_2$ are neighbours in $\mathcal{U}^d(v)$. 
Thus, $\sigma(w_1)$ and $\sigma(w_2)$ must be neighbours in $\mathcal{U}^d(u)$, and so $\pi(x_1) = \text{id}(\sigma(w_1))$ and $\pi(x_2) = \text{id}(\sigma(w_2))$ are neighbours in $G_2$. 

By symmetry of the argument, taking $\{x_1, x_2\} \in \Tilde{E}(u)$ and using the inverses of $\pi$ and $\sigma$, yields that $\pi$ is an isomorphism between $\text{comp}(v)$ and $\text{comp}(u)$. Thus extending $\pi$ by the identity for all nodes in $V(G_1 \cup G_2)\backslash (N^d(v) \cup N^d(u))$ yields an automorphism on $G_1\cup G_2$. 
\end{proof}

Hence, our definition of node roles is rigorous in the sense that for large $d$ it is as expressive as automorphic equivalence, as is well defined locally in the sense that $u\sim_{\mathcal{I}}^d v$ induces a local isomorphism between $G[N^{d-1}(v)]$ and $G[N^{d-1}(u)]$. 
However, the above criterion is hard to check computationally. 
Specifically, a direct consequence of the above proposition is that computing the $d$-roles is at least as hard as solving the graph isomorphism problem. 
The fastest known algorithm for this requires quasi-polynomial time \cite{babai2016graph}, which is intractable for most large problems. 
For fixed $d$, the problem may be computable in polynomial time, but it is still linked to local isomorphism. 
In the following, we therefore propose relaxations of our local role definition that allow efficient computation. 

\section{Relaxations}
\label{sec:computation}

In this section we provide two relaxations of the problem of computing $d$-roles that can be computed efficiently. 
These relaxations are dual in the following sense.
In the first case, we drop the node identifiers in the (identified) unravellings and use node-unidentified unravellings instead.
In the second case, we neglect the detailed knowledge about the (identified) unravelling structure, but keep the information of the node identities at every step.
We start with the first case, which is closely connected to some well known graph-isomorphism algorithms.

\begin{definition}
The \textit{$d$-depth WL node role} ($d$-wl-role) is given by the equivalence induced by the node-unidentified unravellings.
\end{definition}

The equivalence relation $\sim_{\mathcal{U}}^d$ given by this relaxation coincides with the coloring after $d$ iterations of the so-called color refinement algorithm, which is also known as the 1-dimensional Weisfeiler Lehman algorithm \cite{weisfeiler1968reduction}. 
Starting from a constant initial coloring, this algorithm iteratively computes the node colors according to the following formula:
$$ c^{t+1}(v) = \text{hash}\left(c^t(v), \{\!\{c^{t}(x)|x \in N(v)\}\!\}\right) $$
where $\text{hash}$ is an injective hash-function, and $\{\!\{\cdot\}\!\}$ denotes a multiset (a set in which elements can appear more than once).
With every iteration the algorithm aggregates information from its neighbours who, in turn, have aggregated information from their neighbours previously and so on. 
After $d$ iterations, the colors of nodes have information about nodes that are at most distance $d$ apart. 
This information is exactly captured in the node-unidentified unravelling:

\begin{proposition}
\label{prop_color_ref}
Let $G$ be a graph and let $c^d$ be the colors of the color refinement algorithm after $d$ iterations. 
Then the equivalence relation induced by the coloring $c^d$ corresponds to the equivalence relation induced by the $d$-step unravelling $\sim_{c^d} \ \equiv \ \sim_{\mathcal{U}}^d$.
\end{proposition}
\begin{proof}[Proof sketch]
    Consider the color assignment $c^1(v)$ in the first step for a node $v$. 
It encodes the degree of $v$ as the injective hash-function enables the reconstruction of the multi-set of neighboring colors. 
Since all colors are initially the same, this multi-set has one element with muliplicity of $\text{deg}(v)$. 
Following the same logic, in the second iteration the color $c^2(v)$ encodes how many neighbors have which degree. 
Iterating this idea results in the above claim. 
\end{proof}

While not phrased in terms of node roles, the result of Prop. \ref{prop_color_ref} is essentially known in the literature, which is why we only sketch the proof here.
Indeed, color refinement is a well-known algorithm. 
For example, \cite{babai1979canonical} showed that it distinguishes almost all random graphs and there are also results linking the expressive power to fragments of logic \cite{cai1992optimal} and, more recently, graph neural networks \cite{morris2019weisfeiler, xu2018powerful}.
Nonetheless, thinking in terms of $d$-roles provides a fresh look  on both the algorithm and the problem at hand. 
Typically, we care only about the (final) \textit{stable} coloring given by the algorithm, which yields the coarsest equitable partition of the graph. 
However, here we are interested in the preliminary coloring of depth $d$ as it provides us with (relaxed) node roles. 
Moreover, the computation of these $d$-wl-roles is quite efficient: $k$ iterations of color refinement can be computed in time $\mathcal{O}(k \cdot |V(G)| \cdot \text{deg}(G)\cdot \log(\text{deg}(G)))$, where $\text{deg}(G)$ is the maximum degree of $G$. 
While this efficiency is our main motivation to use this algorithm, as our experiments show in the following section, the resulting $d$-wl-roles are very effective for graph analysis tasks as well.

Let us now consider a second, dual relaxation of $d$-roles, where instead of removing the node identifiers, we instead remove the structure in the unravelling. 
More specifically, we count the multiplicity of unique identifiers at every depth of the unravelling and use this data to define the node role. 
The exact information which node is connected to which other node is thus generally neglected in our node assignment.
Still, the expressivity of the assigned node roles is empirically comparable to color refinement.
Moreover, this second procedure performs much better on regular graphs, where color refinement is known to fail, and provides an intuitive and simply explainable representation. 

\begin{definition}
    The \textit{$d$-depth SNP node role} ($d$-snp-role) is given by the multi-set of identifiers in the unravelling at every level up to $d$. 
\end{definition}

From the definition of the node-identified unravelling, we see that counting the identifiers at level $d$ is equivalent to counting the number of walks of length $d$ from the root to all nodes reachable in exactly $d$ steps. 
It is thus closely tied to the $v$-th row  of the adjacency matrix power $(A^d)_{v,\_}$.
However, this row-vector will depend on the ordering of the nodes. 
Instead, we consider the node embedding:
$$\text{SNP}^d(v) =\text{lex-sort}\left( \begin{bmatrix}
A^0_{v, \_} \\
\vdots \\
A^d_{v, \_}
\end{bmatrix}\right)$$
where lex-sort sorts matrix columns lexicographically. 
We call this \textit{sorted Neighbourhood Propagation} (SNP) embedding.
The SNP embedding is isomorphism-invariant, since a permutation of the adjacency matrix preserves the multi-set of columns of the matrix input to the lex-sort and lex-sort, in turn, uniquely determines the location of any column up to equality. 
Computing the embedding takes time $\mathcal{O}(k \cdot |V(G)|^{2.37})$, in general though sparse matrix multiplication speeds up the process for typical graphs. 
We thus have a second relaxation of node-roles based on SNP, where we say two nodes have the same role if their SNP embedding is the same.

\section{Applications/Experiments}
\label{sec:experiments}

The following section presents the results of two computational experiments.
First, we examine the local roles obtained from the WL and SNP algorithm, and see how they can be employed for the analysis of small (188 graphs), medium (600 graphs) and large (2000 graphs) data-sets, namely  MUTAG \cite{debnath1991structure}, ENZYMES \cite{schomburg2004brenda} and NCI1 \cite{wale2008comparison}. 
In all three datasets, nodes are annotated with attributes and there exists a target label for each graph. 
A detailed description of the datasets can be found in \cite{Morris+2020}.
In our second experiment, we compare the results of training a multi-layer perceptron (MLP) classifier using the SNP-embedding as inputs with directly training a graph neural network. 
Throughout both experiments we use node-level and graph-level targets. 
Code will be made available  \href{https://git.rwth-aachen.de/michael.scholkemper/scale-dependent-node-roles}{here}.

\begin{figure}[tb]

\begin{minipage}[b]{1.0\linewidth}
  \centering
  \centerline{\includegraphics[width=8.5cm]{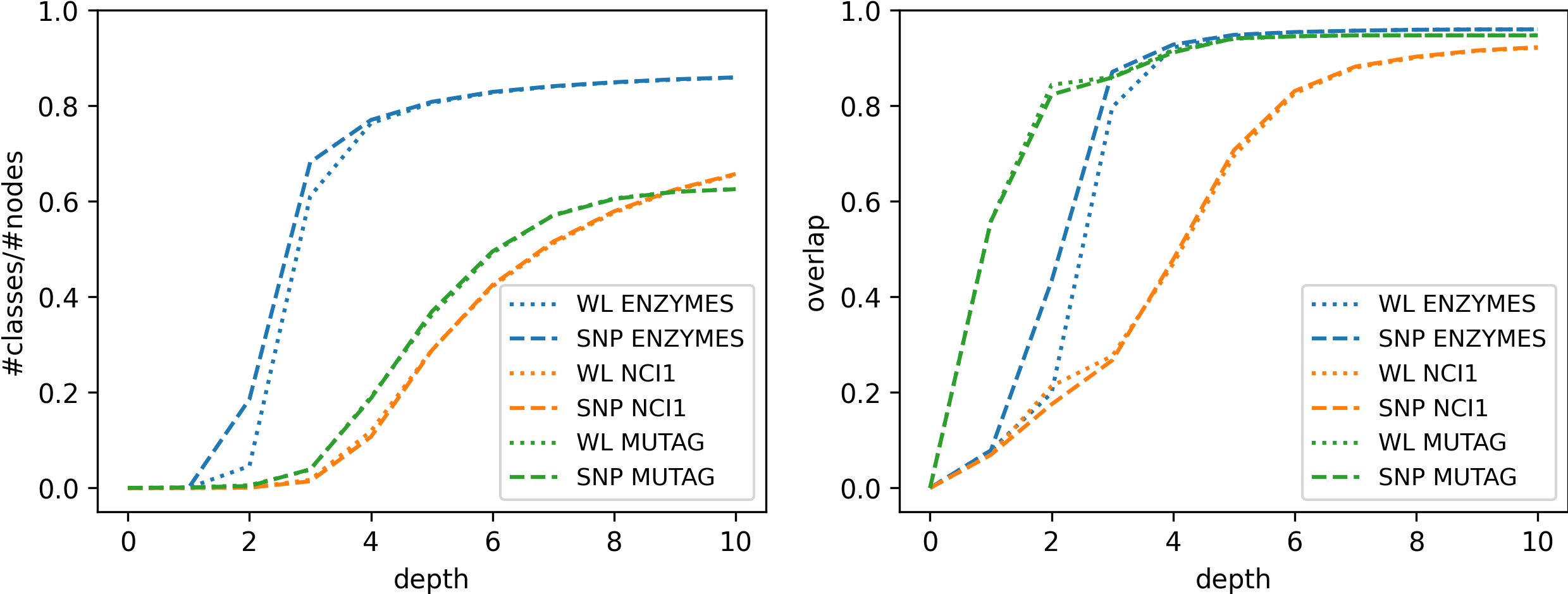}}
\end{minipage}
\caption{\textbf{Disriminative power and utility of local roles.} Number of roles relative to the dataset size (left) and overlap (right) versus depth $d$. Dotted lines indicate the d-wl-role, dashed lines the d-snp-role.}
\label{fig:combined_view}
\end{figure}

\noindent\textbf{Experiment 1.}
We investigate the utility of d-wl-roles and d-snp-roles for a node classification task. 
To this end, Figure \ref{fig:combined_view} shows the number of distinct color classes/embeddings that both algorithms find at each depth $d$, divided by the number of nodes in the datasets.
As can be seen, the local roles are highly discriminitative, already at small depths.
To determine whether these local node roles are useful, we check whether the node roles classes are correlated with some of the node attributes and find that this is indeed the case.
Figure \ref{fig:combined_view} shows the overlap score obtained, if we simply matched the nodes within a role-class with the most frequently occurring label.
The overlap score here is defined as ${(a-b)/(1-b)}$, where $a$ is the accuracy of the assignment of node roles to the node label, and $b$ is a baseline accuracy. 
We use the accuracy at depth $0$ as the baseline, i.e., the accuracy of simply always guessing the most probable label.

Remarkably, a depth of 3--4 suffices to obtain an overlap score of more than $90 \%$ for the three data-sets except NCI1. The latter requires a neighborhood depth of $7$ or more to achieve a similar score. 
Obviously, both of our role-detection methods are not directly suitable as node label classifiers, as they do not generalize to unseen data.
Rather, our results show that the information needed to obtain high accuracy scores in the here considered node classification tasks is present in the near surroundings of the nodes --- without specifying how to exploit it. 

\noindent\textbf{Experiment 2.}
In the second experiment, we address this aspect by comparing the GIN graph neural network architecture \cite{xu2018powerful} --- provably one of the most powerful GNNs, while still permutation invariant --- with a multi-layer-perceptron (MLP) classifier that is given the SNP embedding as input. The GIN uses a 2-layer MLP to update the embeddings and a 3-layer MLP as the final classification layer. For comparability, the final classifier in the SNP approach has the same size. Table \ref{tab:results2} shows the mean accuracy on the test sets for 10 separate $10$-fold cross-validations for each model. In the hyperparameter search, the SNP classifier was only allowed a depth of $3-4$, whereas the GIN was allowed a depth of up to $10$.

\begin{table}[tb]
\small
\begin{tabular}{c c c c c}

& & MUTAG & ENZYMES & NCI1\\ 
\toprule
\multirow{2}{*}{Node-Level} &SNP-   & $\bm{96.7\pm0.1}$ & $\bm{65.8\pm0.1}$ &$\bm{86.4\pm0.0}$\\
 &GIN-  & $96.1\pm0.1$ & $58.4\pm0.2$ &$\bm{86.7\pm0.1}$\\ \midrule
 \multirow{3}{*}{Graph-Level}&SNP-  & $\bm{96.5\pm0.5}$ & $51.0\pm0.6$ &$79.3\pm0.1$\\
 &GIN- & $94.3\pm0.3$ & $31.8\pm0.7$ &$75.6\pm0.2$\\
 &GIN+ & $94.3\pm0.5$ & $\bm{60.3\pm0.7}$ &$\bm{82.0\pm0.3}$\\
\end{tabular}
\caption{\textbf{Overlap (in $\%$) of the SNP classifier and the GIN on the node-level and the graph-level tasks.} {$+$ and $-$ indicate whether the model had access to the node attributes or not. We report the mean over 10 crossvalidation runs along with the standard deviation.}}
\label{tab:results2}
\end{table}

\noindent All in all, the SNP classifier is competitive with the GIN on the node-level and on the graph-level. The comparison with the uninformed GIN- shows that the SNP classifier has the edge when it comes to extracting information --- even if the GIN was significantly deeper. Interestingly, the SNP classifier tended to overfit --- reaching $>99\%$ accuracy on nearly all training datasets --- whereas the GIN tended to achieve similar accuracy on the train-splits as on the test-splits.

\section{Conclusion}
\label{ssec:conclusion}

We formalized the idea of scale-dependent node roles, presented two algorithms to compute such roles and demonstrated their representational power for certain graph learning tasks.
Many recent developments in machine-learning see a strive toward deep classifiers. 
However, certain architectures, such as the GCN architecture \cite{kipf2016semi}, are not suited to be very deep ~\cite{cai2020note, oono2019graph}. 
Our experiments indicate that some graph learning tasks may indeed not require deep features (at least in terms of role depth) and that simple classifiers based on local node roles can yield surprisingly competitive results.
Possible future directions include work on efficient algorithms to compute $d$-roles, incorporating external node features into the SNP-embedding, or establishing other similarity scores based on these ideas.


\vfill\pagebreak

\bibliographystyle{IEEEbib}
\bibliography{refs}

\end{document}